\documentclass[superscriptaddress,twocolumn,floatfix,aps]{revtex4-2}
\usepackage[utf8]{inputenc}
\usepackage{xcolor}
\usepackage{booktabs}
\usepackage{bbm}
\usepackage{bm}
\usepackage{physics}
\usepackage{graphicx}
\usepackage{placeins}
\usepackage{float}
\usepackage{chronology}
\usepackage{amsmath,amsfonts,amssymb,amsthm}
\usepackage{qcircuit}
\usepackage{indentfirst}
\usepackage{dsfont}
\usepackage[normalem]{ulem}
\usepackage{mathtools}
\usepackage{caption}
\usepackage{subcaption}
\usepackage{hyperref}
\usepackage{tcolorbox}
\usepackage{array}
\newcommand{\rvline}{\hspace*{-\arraycolsep}\vline\hspace*{-\arraycolsep}}

\begin{document}

\title{Variational quantum state discriminator for supervised machine learning}

\author{Dongkeun Lee}
\affiliation{Department of Chemistry, Sungkyunkwan University, Suwon, 16419, Republic of Korea}

\author{Kyunghyun Baek}

\affiliation{Electronics and Telecommunications Research Institute, Daejeon, 34129, Republic of Korea}

\author{Joonsuk Huh}
\email{joonsukhuh@gmail.com}
\affiliation{Department of Chemistry, Sungkyunkwan University, Suwon, 16419, Republic of Korea}
\affiliation{Sungkyunkwan University Advanced Institute of Nanotechnology, Suwon, 16419, Republic of Korea}
\affiliation{Institute of Quantum Biophysics, Sungkyunkwan University, Suwon, 16419, Republic of Korea}

\author{Daniel K. Park}
\email{dkd.park@yonsei.ac.kr}
\affiliation{Department of Applied Statistics, Yonsei University, Seoul, 03722, Republic of Korea}
\affiliation{Department of Statistics and Data Science, Yonsei University, Seoul, 03722, Republic of Korea}

\begin{abstract}
Quantum state discrimination (QSD) is a fundamental task in quantum information processing with numerous applications. We present a variational quantum algorithm that performs the minimum-error QSD, called the variational quantum state discriminator (VQSD). The VQSD uses a parameterized quantum circuit that is trained by minimizing a cost function derived from the QSD, and finds the optimal positive-operator valued measure (POVM) for distinguishing target quantum states. The VQSD is capable of discriminating even unknown states, eliminating the need for expensive quantum state tomography. Our numerical simulations and comparisons with semidefinite programming demonstrate the effectiveness of the VQSD in finding optimal POVMs for minimum-error QSD of both pure and mixed states. In addition, the VQSD can be utilized as a supervised machine learning algorithm for multi-class classification. The area under the receiver operating characteristic curve obtained in numerical simulations with the Iris flower dataset ranges from 0.97 to 1 with an average of 0.985, demonstrating excellent performance of the VQSD classifier.
\end{abstract}

\maketitle

\section{Introduction}
\label{sec:introduction}
Quantum measurement theory plays a crucial role in quantum information processing (QIP), driving advancements in communication, computation, and sensing~\cite{PhysRevLett.68.3121,PhysRevLett.80.4999,doi:10.1080/00107510010002599,PhysRevX.9.041029}.
The theory states that non-orthogonal quantum states can be distinguished with non-zero probability through quantum measurement, which can generally be described by Positive Operator-Valued Measures (POVMs)~\cite{NC00,watrous_2018}.
This arises from the geometric structure of quantum states defined on a Hilbert space and the measurement postulate of quantum mechanics. Quantum state discrimination (QSD) is a well-established field that provides a theoretical ground for the distinguishability of quantum states and the retrieval of classical information. Notably, the optimal strategy for distinguishing two quantum states was proposed even prior to the advent of quantum computing~\cite{Helstrom1969}.

The ability to distinguish non-orthogonal states offers exciting prospects for data science and machine learning as an arbitrary number of data can be encoded in a single qubit. There is no classical analog to this since a classical bit can only represent two orthogonal states. As a result, the concept of QSD has emerged in various contexts within machine learning. One area of research focuses on utilizing QSD for classical-quantum hybrid machine learning. For instance, the optimal measurement theory for two-element POVMs has been applied to determine the best quantum feature map for binary classification tasks~\cite{lloyd2020quantum} and to learn a quantum circuit for quantum data classification with a limited set of two-qubit states~\cite{chen_qumi2020,PhysRevResearch.3.013063}. Several quantum-inspired algorithms~\footnote{Quantum-inspired algorithms refer to algorithms executed on classical hardware but using the mathematical formalism of quantum mechanics.} based on the theory of QSD for constructing classifiers have also been reported~\cite{cagliari.hqc,giuntini2021quantum,GIUNTINI2023109956}. 
However, finding the optimal measurement for QSD using a classical computer becomes computationally intractable for a large number of qubits, as it necessitates complete information about the states, often obtained through quantum state tomography. In addition, the two-state QSD requires a spectral decomposition to construct the optimal POVM. Furthermore, the optimization for more than two states is typically performed through semidefinite programming (SDP), which involves a number of steps that increase polynomially with the dimension of the Hilbert space and thus, exponentially with the number of qubits~\cite{PhysRevResearch.3.013063}. Furthermore, converting the optimal POVM found from classical methods to a corresponding quantum circuit for implementation is generally a challenging task. Given these challenges, it is natural to explore the possibility of directly finding the optimal POVM for QSD on a quantum computer.

Motivated by the broad applications of QSD in various fields, including machine learning, and the challenges of classical optimization, we present a variational quantum algorithm (VQA) for performing minimum-error QSD. Our VQA is a general solution capable of discriminating quantum systems of any dimension, whether pure or mixed, in a systematic manner. Based on the variational quantum state discriminator (VQSD) for mixed states, we also propose a quantum multi-class classification algorithm. The training of the quantum circuit corresponds to identifying non-linear decision boundaries within the Hilbert space where the data is embedded, as the number of distinguishable classes can exceed the dimension of the Hilbert space. While this work primarily demonstrates the applicability of the VQSD in machine learning, it can be applied to any application requiring minimum-error QSD.

The remainder of the paper is organized as follows. In Sec.~\ref{sec:theoretical}, we review theoretical backgrounds for our work, such as the quantum circuit for POVMs and the minimum-error QSD. Therein, we generalize a previous work that reported the gate decomposition for four-element POVMs on two-qubit states to an arbitrary number of qubits and POVM elements. Section~\ref{sec:vqsd} presents the VQSD. It describes how to construct the cost function and the training procedure in detail. The numerical simulation and results are also described at the end of the section, demonstrating the success of VQSD in finding the optimal POVM. Section~\ref{sec:ml} describes the application of the VQSD in supervised machine learning and show simulation results for multi-class classification with the Iris flower dataset. Conclusions, discussion, and potential future research directions are provided in Sec.~\ref{sec:conclusion}.

\section{Theoretical Framework}
\label{sec:theoretical}
\subsection{Quantum Circuit for POVMs}

\begin{figure*}[t!]
  \centering
  \begin{minipage}[t]{.55\textwidth}
    \subcaptionbox{\small ($n_T+n_A$)-qubit unitary circuit $\mathcal{U}$}
      {\includegraphics[width=\textwidth]{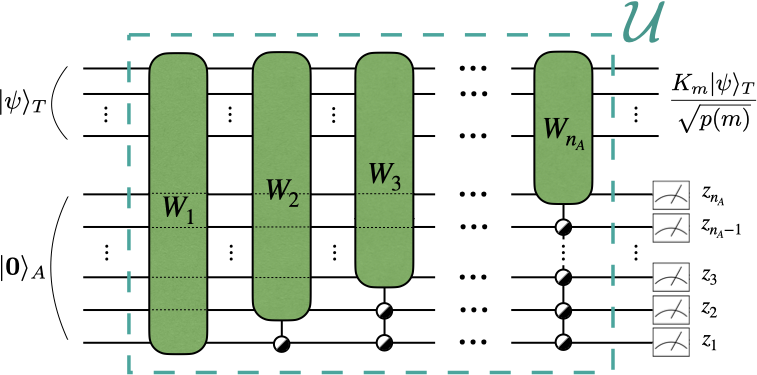}}%
  \end{minipage}%
  \begin{minipage}[b]{.45\textwidth}
    \subcaptionbox{\small ($n_T+1$)-qubit gate $W_a$}
      {\includegraphics[width=.55\textwidth]{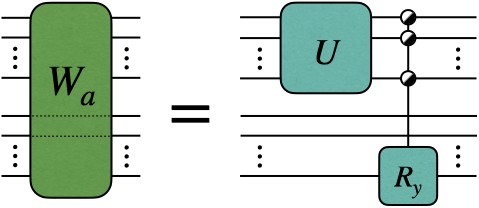}}
    \\
    \subcaptionbox{\small CSD binary tree}
      {\includegraphics[width=.65\textwidth]{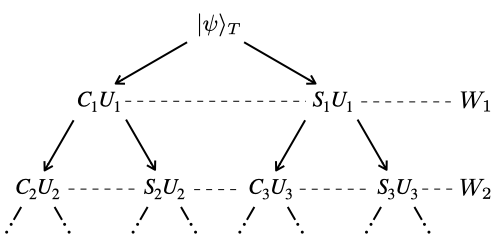}}%
  \end{minipage}%
  \caption{\small The circuit diagram for the unitary operation $\mathcal{U}$. It describes performing POVMs on the arbitrary quantum states or target states $|\psi\rangle_T$. Which POVM element is performed is determined by measuring ancillary qubits and the bit string outcomes $z_1z_2\cdots z_{n_A}$ correspond to the POVM outcomes $m$. (a) A circuit $\mathcal{U}$ is factorized into  $n_A$ numbers of (uniformly controlled) $W_a$ gates and (b) a ($n_T+1$)-qubit gate $W_a$ consists of a general $n_T$-qubit $U$ and an uniformly controlled $R_y$. The dotted lines drawn over $W_a$ gates indicate that the corresponding qubits are not involved in the operation and the half-filled circle stands for the uniform controls. (c) The CSD tree describes binary decisions generated from ancillary qubits of the circuit $\mathcal{U}$ at a glance.}
  \label{fig:circ_decomp}%
\end{figure*}

A measurement in quantum mechanics 
is in general described by a set of positive semi-definite operators $\lbrace E_m\rbrace$ satisfying $\sum_{m=0}^{l-1} E_m = I$. Here, $m\in\lbrace 0,\ldots,l\rbrace$ labels the possible measurement outcomes, and the probability to obtain $m$ given a density matrix $\rho$ that describes the quantum state is given by $p_m = \Tr(E_m \rho)$. Note that the operators can be decomposed as $E_m = K_m^{\dagger}K_m$ where $\lbrace K_m\rbrace$ is known as a set of the Kraus operators. Upon obtaining the outcome $m$, the quantum state evolves to $K_m\rho K_m^{\dagger}/p_m$. We refer to such measurement procedure with the collection of matrices $E_m$ as an $l$-element Positive Operator-Valued {Measure} (POVM) \cite{NC00}.

According to Neumark’s theorem, an arbitrary
POVM can be implemented with a quantum circuit by applying a unitary operator on an extended system consisting of the target quantum state and an ancillary system and then measuring the latter~\cite{neumark.1976,peres.1990}.
In this regard, ancillary qubits are introduced and a multi-qubit unitary gate denoted by $\mathcal{U}$ acts on all qubits.  
The unitary operation $\mathcal{U}$ on target and ancillary qubits can be described in terms of a set of the Kraus operators $\{K_m\}$ as
\begin{equation}\label{eq:kraus_circ}
    \mathcal{U}|\psi\rangle_T|\mathbf{0}\rangle_A = \sum^{l-1}_{m=0} \left(K_m|\psi\rangle_T\right)|m\rangle_A,
\end{equation}
where $|\psi\rangle_T$ is the target quantum state consisting of $n_T$ qubits and $|\mathbf{0}\rangle_A$ denotes all ancillary qubits in the state $|0\rangle$ (See Fig.~\ref{fig:circ_decomp}(a)).
Equation~\eqref{eq:kraus_circ} signifies the implementation of the quantum channel of Kraus rank $l$ \cite{Iten.2017, Yordanov.2019}.
Performing a POVM on $|\psi\rangle_T$ is then embodied by measuring ancillary qubits in the computational basis (e.g. projective measurement in the $Z$ basis), which yields the outcome $m$ with the probability $p(m)={}_T\langle\psi|K^\dagger_mK_m|\psi\rangle_T$. Here, we call the unitary operation $\mathcal{U}$ the POVM circuit.

Since an arbitrary unitary gate can be parameterized with real-valued parameters, the POVM operator that produces the outcome $m$ can be written as
\begin{align}\label{eq:povm_circ}
    \nonumber \mathcal{E}_m(\Theta) &= E_m(\Theta) \otimes |\mathbf{0}\rangle_A\langle\mathbf{0}| \\
    & = \mathcal{U}^\dagger(\Theta) (I_T\otimes M_m)\,\mathcal{U}(\Theta),
\end{align}
where $\Theta$ is the parameter vector, $I_T$ is the identity operator acting on the target qubits, and $M_m=|m\rangle_A\langle m|$ represents the projective measurement on the ancillary qubits in the computational basis. Note that each bit string $z_1z_2\cdots z_{n_A}\in\lbrace 0,1\rbrace^{n_A}$ produced by measuring $n_A$ ancillary qubits is converted into the decimal value $m\in\{0,\cdots, l-1\}$, the outcome index of POVMs.

There are several ways to decompose the multi-qubit quantum circuits such as the cosine-sine decomposition \cite{Tucci.1999,Shende.2006}, the decomposition introduced by Knill \cite{Knill.1995}, and the column-by-column decomposition~\cite{Iten.2016}. 
For a better understanding of the implementation of POVMs in the quantum circuit, we opt for a proper unitary decomposition, the diagrams of which are illustrated in Fig.~\ref{fig:circ_decomp}. The decomposition of $\mathcal{U}$ for $n_T=n_A=2$ is discussed in Ref.~\cite{chen_qumi2020}. We generalize the result therein for any $n_T$ and $n_A$ based on the cosine-sine decomposition (CSD) and provide its rigorous development in Appendix~\ref{append:decomp}. Specifically, an arbitrary quantum circuit $\mathcal{U}$ is first decomposed into a set of ($n_T+1$)-qubit unitary operators $W_a$, uniformly controlled by $a-1$ ancilla qubits as depicted in Fig.~\ref{fig:circ_decomp}(a).
A unitary operator $W_a$ is then decomposed into a general $n_T$-qubit unitary gate $U$ and an uniformly controlled single-qubit rotation around the $y$-axis of the Bloch sphere, denoted by $R_y$, is applied to the $a$th ancillary qubit controlled by all target qubits as shown in Fig.~\ref{fig:circ_decomp}(b). 
The dimension of the parameter vector $\Theta$ from all $U$ and $R_y$ gates is $(l-1)(2^{2n_T}-1 + 2^{n_T})$.

Let us scrutinize the circuit structures of Fig.~\ref{fig:circ_decomp}(a) and (b) from the perspective of the Kraus operator. The action of the uniformly controlled $R_y$ gate in Fig.~\ref{fig:circ_decomp}(b) is equivalent to the multiplication of a diagonal matrix by a state vector in the target system.
Depending on the state of the ancillary qubit, the diagonal matrix is either $C:=\cos(diag(\Theta_{R}))$ if the ancilary qubit is in the state $|0\rangle_A$, or $S:=\sin(diag(\Theta_R))$ for the state $|1\rangle_A$, where $\Theta_{R}=(\theta_1,\cdots,\theta_{2^{n_T}})$ is a parameter vector involving $2^{n_T}$ angles of $R_y$ gates.
Thus, a single gate $W_a$ implies the implementation of the Kraus operators of rank 2, $\{CU, SU\}$. 

In order to generate the Kraus operator of rank $l>2$, an ancillary system of $n_A = \lceil \log_2 l \rceil $ qubits is required and $(a-1)$-fold uniformly controlled $W_a$ gates are applied consecutively as the number of control qubits are increased up to $n_A-1$, as shown in Fig.~\ref{fig:circ_decomp}(a). 
Each uniformly controlled $W_a$ gate builds up the binary rank Kraus operators that depend on the previous results of the ancillary system.
The whole procedure can be figured out by introducing a CSD binary tree (see Fig.~\ref{fig:circ_decomp}(c)) \cite{Tucci.1999}.
Then the Kraus operator of rank $l$ can be formulated as
\begin{equation}
    K_m = \prod^{n_{A}}_{a=1}D_{j(a)}U_{j(a)},
\end{equation}
where a diagonal matrix $D_{j(a)}$ becomes either $C_{j(a)}$ if $z_{a}=0$, or $S_{j(a)}$ if $z_{a}=1$. 
Here, the index $j(a):=2^{a-1}+\sum^{a-1}_{i=1}z_i 2^{a-1-i}$ is introduced to enumerate the actions of the uniformly controlled $W_a$ gate because it is comprised of a sequence of $(a-1)$-fold controlled gates with different angles (see Fig.~\ref{fig:ucgate} in Appendix~\ref{append:decomp}).
For example, a set of rank-4 Kraus operators, $\{C_2U_2C_1U_1, S_2U_2C_1U_1, C_3U_3S_1U_1, S_3U_3S_1U_1\}$, can be carried by using $W_1$ and uniformly controlled $W_2$ gates and one of its elements is determined by the measurement outcome $z_1z_2$.

The implementation of the control part of the uniformly controlled $W_a$ gates can be performed classically as it commutes with measurements, as detailed in the appendix (Appendix~\ref{append:decomp}). As a result, the quantum gate complexity is primarily determined by the implementation of the $W_a$ gate depicted in Fig.~\ref{fig:circ_decomp}(b).
In the implementation of $W_a$, the complexity of the required number of {\small CNOT} gates in an arbitrary $n_T$-qubit unitary $U$ is $O(2^{2n_T})$ \cite{Shende.2006}. Additionally, the uniformly controlled $R_y$ operation involves $2^{n_T}$ {\small CNOT} gates \cite{Mottonen.2004}.
Thus the total number of {\small CNOT} gates in the circuit $\mathcal{U}$ is $O(2^{2n_T}l)$ for a given value of $l$.
In the case of $n_T=2$, the total number of {\small CNOT} gates reduces to $3(l-1)$, where three is considered the smallest number of {\small CNOT} gates for a universal two-qubit gate \cite{Shende.2004, Vatan.2004, Vidal.2004}.
To mitigate the exponential increase in the number of two-qubit gates required for $\mathcal{U}$, it may be desirable to choose a simpler structure for $\mathcal{U}$ such as the ansatz used in the Variational Quantum Eigensolver (VQE) \cite{PhysRevResearch.3.013063}. However, this simplification comes at the cost of sacrificing global optimality.

\subsection{Minimum-error quantum state discrimination}\label{sec:med}

Quantum state discrimination is a protocol that aims to identify a quantum state among a set of \textit{a priori} completely known candidate states. The POVM framework allows for the discrimination of non-orthogonal states with the optimal success probability which depends on the distance between \textit{a priori} quantum states. This naturally motivates the minimum-error QSD whose goal is to maximize the success probability for correctly guessing a state.

In preparation for minimum-error QSD, let us consider a set of $l$ different quantum states $\{\rho_n\}^{l-1}_{n=0}$ with the corresponding \textit{a priori} probabilities $\{q_n\}_{n=0}^{l-1}$. When POVMs $\{E_m\}$ are performed on quantum states $\rho_n$, the probability of the outcome $m$ is $p(m|n)=\Tr[E_m\rho_n]$. If appropriate POVMs are taken, one can correctly guess a state $\rho_m$ with a high probability $q_mp(m|m)$. In this context, the minimum-error QSD aims to minimize the error probability\
\begin{equation}\label{eq:med}
p_{error} = 1- \sum_{m=0}^{l-1}q_m\Tr[\rho_mE_m],
\end{equation}
where the last term denotes the success probability of the given POVM~\cite{Barnett:09, Bae_2015}.

If only two elements of POVMs ($l=2$) are implemented for discriminating two different quantum states, one can analytically find the optimal POVMs and the associated minimum probability of the error Eq.\eqref{eq:med}, known as the Helstrom bound $\mathcal{B}_H$ in this special case~\cite{Helstrom1969}. Starting from Eq.\eqref{eq:med}, the minimum error probability can be rewritten as
\begin{equation}\label{eq:Helstrom}
    \min p_{error} = \frac{1}{2} - \max_{E}\frac{1}{2}\Tr[E\Lambda] 
\end{equation}
where $E=E_0-E_1$ and $\Lambda = q_0\rho_0-q_1\rho_1$. By applying the spectral decomposition, one can get $\Lambda=\lambda_+|\lambda_+\rangle\langle\lambda_+| + \lambda_-|\lambda_-\rangle\langle\lambda_-|$ where $\lambda_+$ and $\lambda_-$ denote positive and negative eigenvalues of $\Lambda$, repsectively. 
The optimal POVMs can be obtained as $E_0=|\lambda_+\rangle\langle \lambda_+|$ and $E_1=|\lambda_-\rangle\langle\lambda_-|$. 
The Helstrom bound is then given by $\mathcal{B}_H := \frac{1}{2}-\frac{1}{2}\Tr|\Lambda|$.

When $l$ is larger than two, the set of POVMs that minimizes Eq.\eqref{eq:med} usually does not have an analytical solution. Instead, the necessary and sufficient conditions are known to be satisfied by a set of the optimal POVMs and are expressed as
\begin{align}
    & E_m\left(q_m\rho_m-q_n\rho_n\right)E_n = 0 \\
    & \sum^{l-1}_{m=0} q_m E_m\rho_m - q_n\rho_n \ge 0
\end{align}
for any $m, n = 0,\cdots,l-1$ \cite{holevo1973statistical,Helstrom1969,yuen1975optimum}. 
It is noted that these conditions are equivalent to each other and closely related to the duality relationship of the semidefinite program \cite{yuen1975optimum}.
The minimum-error QSD problem can be formulated as the SDP.
The classical solver for the SDP, however, requires the number of iterations that grows exponentially with the number of target qubits and polynomially with the number of the POVM elements as $O(2^{\omega n_T}l^{\omega})$, where $\omega$ is the constant ranges from 2.37188 to 3 depending on the algorithm for matrix multiplication (see Appendix \ref{append:sdp}). In addition, the objective function involves computing the outcome probabilities of POVM, and hence computing it classically requires exponential runtime.
The aforementioned procedure assumes that the target density matrices are fully known. If not, quantum state tomography, which adds another layer of exponential complexity, is also required.
The intractability of minimum-error QSD via classical means motivates the development of quantum algorithms for it. Our method, which is described in the following section, is based on training a parameterized quantum circuit, aiming to realize it on the noisy intermediate-scale quantum (NISQ) devices~\cite{Preskill2018quantumcomputing}. Later, we will use the classical solver for the SDP to compare its result with that of the VQSD \cite{diamond2016cvxpy}.

\section{Variational Quantum State Discrimination}
\label{sec:vqsd}

Training a parameterized quantum circuit with classical optimization algorithms (e.g. gradient-based) have gained much attention recently, especially as an effective means to utilize NISQ devices. Without loss of generality, we refer to an algorithm based on such an approach variational quantum algorithm (VQA)~\cite{cerezo2020variational,bharti2021noisy}.
Here, we develop a VQA that learns to perform the minimum-error QSD in a supervised manner. In other words, the VQA trains a quantum circuit to perform the optimal POVM for a given set of labelled quantum states. 
We call this VQA the variational quantum state discriminator (VQSD).
While the labels for the quantum states need to be known a priori, the quantum states themselves can be completely unknown, and this distinguishes the proposed algorithm from the existing QSD methods. 

\subsection{Cost function}

The first step towards utilizing the VQA for minimum-error QSD is establishing an appropriate cost function that can discriminate different labeled sets beyond individual quantum states.
Let us consider that a set of $N$ quantum states with $l$ labels, 
\begin{equation}
\label{eq:data}
\mathcal{D}=\{(|\Psi_n\rangle,y_n)|\, y_n\in\{0,\cdots,l-1\},\; 
1\le n\le N\},
\end{equation}
is partitioned into $l$ disjoint subsets. 
A subset of quantum states with the same label $m$ can be denoted by $\mathcal{D}_m = \{(|\Psi_n\rangle,y_n)| y_n=m,\,\forall n\in\mathcal{N}_m\}$ with the index set $\mathcal{N}_m=\{n|y_n=m,\,\forall n\in\mathbb{N} \}$.

To discriminate between different subsets $\mathcal{D}_m$ of input states $\{|\Psi_n\rangle_I\}_n$, one can perform POVMs on either all $n_I$ input qubits or $n_T$ qubits out of $n_I$ qubits for each input state $|\Psi_n\rangle_I$.
The latter can be considered as measuring the reduced density matrix of $|\Psi_n\rangle_I$ on the target subsystem consisting of $n_T$ target qubits. 
This reduced density matrix can be written as $\rho_{T,n}=\text{Tr}_{\bar{T}}|\Psi_n\rangle_I\langle\Psi_n|$, where the complement of the target subsystem $\bar{T}$ is traced out.

For given $|\Psi_n\rangle_I$, a POVM $E_m$ may be performed on the target subsystem, which yields the probability of the outcome $p(m|n)=\text{Tr}[E_m \rho_{T,n}]=\,_{I}\langle\Psi_n|E_m|\Psi_n\rangle_I$. 
For given $\mathcal{D}_m$, it is possible to perform a POVM with the outcome $m$ on quantum states in the same class and obtain the probability of the outcome $m$, $p(m|\mathcal{D}_m):=\sum_{n\in\mathcal{N}_m}\,_I\langle\Psi_n|E_m|\Psi_n\rangle_I/|\mathcal{N}_m|$, where $|\mathcal{N}_m|$ is the cardinality of $\mathcal{N}_m$.
This probability can also be viewed as performing $E_m$ on the target subsystem of the mixed state 
\begin{equation}
\label{eq:datamixed}
\rho'_{m} = \sum_{n\in\mathcal{N}_m}|\Psi_n\rangle\langle\Psi_n |/|\mathcal{N}_m|,
\end{equation}
which is the uniform mixture of all states in a subset $\mathcal{D}_m$.

As mentioned in the previous section, one can make POVMs on the input states in the quantum circuit.
From the MED Eq.\eqref{eq:med}, the cost function can then be expressed in terms of the parameterized quantum circuit as
\begin{align}
\nonumber C(\Theta) &= 1-\sum_{m=0}^{l-1} q_m p(m|\mathcal{D}_m) \\
\label{eq:costfn} &= 1-\sum^{l-1}_{m=0}\sum_{n\in\mathcal{N}_m}\frac{1}{|\mathcal{D}|}\langle\Psi^0_n| \mathcal{E}_m(\Theta) |\Psi^0_n\rangle,
\end{align}
where $|\Psi^0_n\rangle = |\Psi_n\rangle_I|\mathbf{0}\rangle_A$ is a quantum state of all qubits and $q_m=|\mathcal{N}_m|/|\mathcal{D}|$ is an \textit{a priori} probability for the subset $\mathcal{D}_m$. 
Minimizing this cost function reduces the error of the label discrimination. This signifies that the VQSD is well-trained from a set of the quantum states and can predict the label of a new quantum state based on the measurement that optimally separates the subsets $\mathcal{D}_m$ according to their label $m$.
Note that the cost function $C(\Theta)$ Eq.\eqref{eq:costfn} is identical to the error probability $p_{error}$ in Eq.\eqref{eq:med} when $|\mathcal{N}_m|=1$ for all $m$.

\subsection{Algorithm}
\begin{figure}[t]
    \centering
    \includegraphics[width=.48\textwidth]{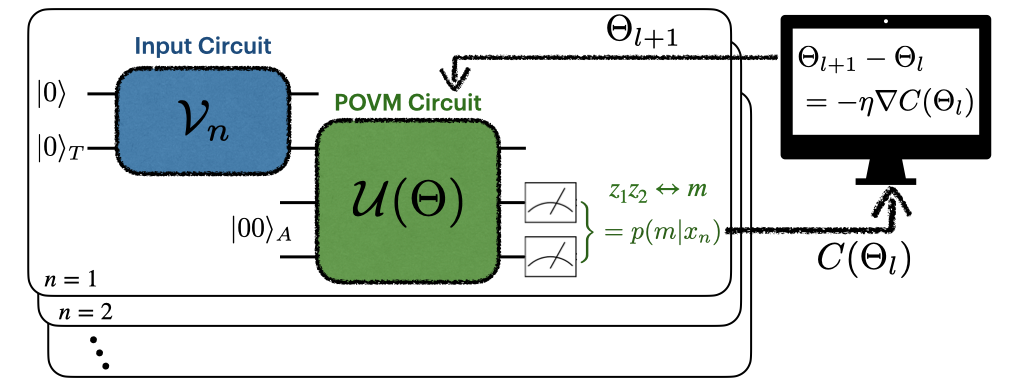}
    \caption{\small Illustration of the variational quantum state discriminator. The input circuit prepares a set of arbitrary quantum states, either pure or mixed, subject to POVM. In this example, the set of two-qubit states is $\{\mathcal{V}_n|00\rangle\}_n$. A parameterized POVM circuit is then performed on the target qubit of the input state ($n_T=1$ and $n_I=2$ in this example) and trained by optimizing the cost function $C(\Theta)$  given in Eq.~\eqref{eq:costfn}} 
    \label{fig:training_circ}
\end{figure}

The VQSD algorithm can be constructed as follows. First, a set of the input states $\{|\Psi_n\rangle\}_{n\le N}$ is given and ancillary qubits are prepared in the state $|\mathbf{0}\rangle_A$. Note that the algorithm only requires the label information and the input quantum states does not need to be known. Hence it bypasses the quantum state tomography. Second, the parameterized POVM circuit $\mathcal{U}(\Theta)$ acts on $n_T$ target qubits out of $n_I$ input qubits, which implies that POVMs can be performed on the reduced density matrix $\rho_{T,n}$ of an input state $|\Psi_n\rangle$. When implementing on NISQ devices, small $n_T$ is desirable. Note that the number of outcomes obtained from POVMs can be controlled by adjusting the number of ancillary qubits.
Third, the outcomes $m$ and the corresponding probabilities $p(m|n)$ of POVMs are obtained from measuring ancillary qubits in the computational basis. 
Fourth, a cost function Eq.\eqref{eq:costfn} is constructed from the outcomes $m$ and probabilities $p(m|n)$ and minimized using classical optimizer. 
The parameter vector $\Theta$ in the POVM circuit is then optimized toward synchronizing the outcome of POVMs with the label of the input states.
The trained POVM circuit $\mathcal{U}(\Theta_{opt})$ with the optimal parameter vector $\Theta_{opt}$ serves as a quantum state discriminator.
In consequence, an unlabeled input state being inserted, the trained POVM circuit and the measurement on ancillary qubits predicts a label with the highest probability. 
The VQSD algorithm is illustrated in Fig.~\ref{fig:training_circ}.

\subsection{Simulation and Results}

We demonstrate the performance of the VQSD via classical simulations. The simulations are carried out using PennyLane~\cite{bergholm2018pennylane}, and the quantum circuit parameters are updated via gradient descent optimization using adaptive moment (ADAM) estimation~\cite{kingma2014adam}.
 
\begin{figure}[t!]
    \centering
    \includegraphics[width=0.35\textwidth]{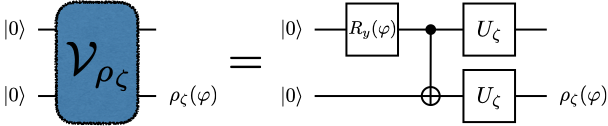}\\
    \caption{\small Circuit diagrams for input-state preparations. (a) A mixed state $\rho_{\zeta}$ is realized by exploiting one of two qubits in the state $\mathcal{V}_{\rho_\zeta}|00\rangle$. The gate $U_{\zeta}$ corresponds to one of three $U_x=H$, $U_y=HS$, and $U_z=I$.}
    \label{fig:state_prep}
\end{figure}

The VQSD is tested with both pure and mixed states. The quantum circuit creates a mixed state by tracing out a subsystem from an entangled state. More specifically, the simulation uses the quantum circuit shown in Fig.~\ref{fig:state_prep}, which prepares an entangled state by applying $\mathcal{V}_{\rho_{\zeta}}$ to $|00\rangle$. The reduced density matrix on the second qubit, denoted by $\rho_\zeta$, is the mixed state inserted as the input to the POVM circuit.
The mathematical form of a single-qubit mixed state $\rho_\zeta$ is then given by
\begin{align}\label{eq:mixedstate}
\rho_\zeta(\varphi) \equiv \cos^2\varphi|\zeta_+\rangle\langle\zeta_+| + \sin^2\varphi|\zeta_-\rangle\langle \zeta_-|,
\end{align}
where $|\zeta_{\pm}\rangle$ denotes
the eigenstates of the Pauli matrices $\sigma_\zeta$ with $\zeta\in\{x, y, z\}$ and $\varphi$ parameterizes the mixture weights.
The Bloch vector of $\rho_\zeta(\varphi)$ lies along the $\zeta$ axis of the Bloch sphere. 
In order to prepare a mixed state $\rho_\zeta(\varphi)$ as an input in the quantum circuit, we make use of a two-qubit unitary gate $\mathcal{V}_{\rho_\zeta}$ illustrated in Fig.~\ref{fig:state_prep} that comprises $R_y$, {\small CNOT}, Hadamard, and phase shift gates.
$R_y(\varphi)$ and {\small CNOT} gates generate an entangled state $\cos\varphi|00\rangle+\sin\varphi|11\rangle$. The $U_{\zeta}$ gate for each qubit changes the $z$ basis into the $\zeta$ basis, $\cos\varphi|\zeta_{+}\zeta_{+}\rangle+\sin\varphi|\zeta_{-}\zeta_{-}\rangle$, where $U_z=I$, $U_x=H$, and $U_y=HS$.
We then proceed with only one qubit after the unitary operation $\mathcal{V}_{\rho_\zeta}$, which signifies that the reduced density matrix of the second qubit, $\rho_2=\text{Tr}_1[\mathcal{V}_{\rho_\zeta}|00\rangle\langle 00|\mathcal{V}^\dagger_{\rho_\zeta}]$, is subject to QSD.

\begin{figure}[t]
\centering
\begin{subfigure}{0.235\textwidth}
    \includegraphics[width=\textwidth]{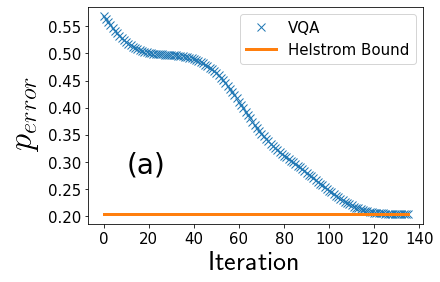}
\end{subfigure}
\hfill
\begin{subfigure}{0.235\textwidth}
    \includegraphics[width=\textwidth]{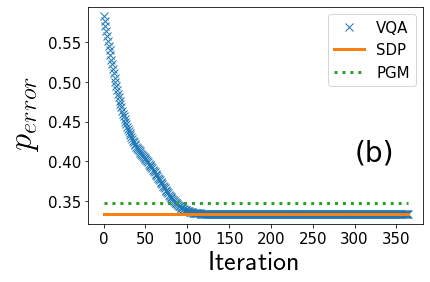}
\end{subfigure}
\\
\begin{subfigure}{0.235\textwidth}
    \includegraphics[width=\textwidth]{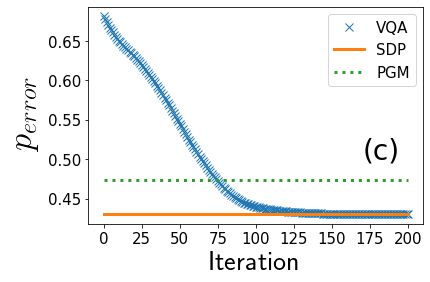}
\end{subfigure}
\hfill
\begin{subfigure}{0.235\textwidth}
    \includegraphics[width=\textwidth]{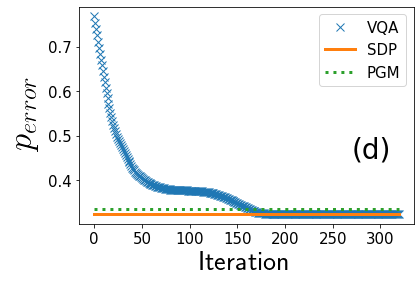}
\end{subfigure}
\caption{{\small Minimum-error QSD using the variational quantum algorithm. Sets of \textit{a priori} quantum states are given as (a) $\{\rho_z(\pi/5), \rho_x(\pi/6)\}$, (b) $\{|0\rangle,|1\rangle,|+\rangle\}$, (c) $\{\rho_z(\pi/5), \rho_x(\pi/6), \rho_y(\pi/8)\}$, and (d) $\{|00\rangle, |{++}\rangle, \frac{1}{\sqrt{2}}(|00\rangle+|11\rangle), \frac{1}{\sqrt{2}}(|01\rangle+|10\rangle)\}$ and \textit{a priori} probabilities are assumed to be equal. The two-state discrimination task in (a) is compared with the Helstrom measurement, and the three-state discrimination tasks in (b) to (d) are compared with the results from the SDP solver and the PGM. Each plot shows that the cost function converges to its minimum after approximately 100 to 150 iterations.}}
\label{fig:med}
\end{figure}

%
Together with the quantum circuit Fig.~\ref{fig:state_prep} of the mixed state Eq.\eqref{eq:mixedstate}, we prepare four \textit{a priori} sets of  quantum states with equal probabilities: 
$\{\rho_z(\pi/5), \rho_x(\pi/6)\}$, $\{|0\rangle,|1\rangle,|+\rangle\}$, $\{\rho_z(\pi/5), \rho_x(\pi/6), \rho_y(\pi/8)\}$, and $\{\frac{1}{\sqrt{2}}(|00\rangle+|11\rangle), \frac{1}{\sqrt{2}}(|01\rangle+|10\rangle), |00\rangle, |{++}\rangle \}$, 
We then construct the parameterized POVM circuit Eq.\eqref{eq:povm_circ} and the cost function Eq.\eqref{eq:costfn} with the condition $m=n$.
By implementing the VQSD, we attain the minimum value of the cost function with the convergence tolerance of around $10^{-7}$. The minimum error of which is depicted in Fig.~\ref{fig:med}.

To verify the correctness of the solution obtained from the VQSD, we compare the two-state discrimination task with the Helstrom measurement, and the three-state discrimination tasks with the semidefinite programming (SDP) solved by a convex optimization tool \cite{diamond2016cvxpy}.
The difference between two minimum errors from the classical SDP solver and the VQSD is around $10^{-6}$, which confirms the validity of the VQSD.
We also compare our method with the pretty-good measurement (PGM) defined as $\Pi_m:=q_m\rho^{-1/2}\rho_m\rho^{-1/2}$ with the ensemble $\rho = \sum_{n=0}^{l-1} q_n\rho_n$, which is a suboptimal $l$-element POVM when $l>2$~\cite{Hughston:1993,Hausladen:1994,Ban:1997}. This measurement entirely depends on the state preparation and is prescribed to gather the information of a quantum system in a simple way with an analytical expression, although the global optimality is not guaranteed. The error probability from performing the PGMs on each set of quantum states is shown in Fig.~\ref{fig:med}(b)-(d), which corresponds to the fact that the optimal POVM is generally not a PGM for the minimum-error QSD task \cite{Andersson.2002}. Simulation results verify that the VQSD finds better measurements than the PGM in all instances.



\section{Application in Machine Learning}
\label{sec:ml}

\subsection{Connection to classification}

We now apply the VQSD to classification, a ubiquitous problem in data science that can be solved via supervised learning. The training dataset for $l$-class classification is the same as $\mathcal{D}$ in Eq.~(\ref{eq:data}), except $|\Psi_n\rangle$ represents the data embedded as the quantum state.
With the training dataset, the VQSD can be trained to perform minimum-error QSD to distinguish $\rho'_m$ in Eq.~(\ref{eq:datamixed}) with $m\in\lbrace 0,\ldots,l-1\rbrace$. Once training is completed, a test datum embedded as a quantum state is fed into the VQSD circuit, and the resulting POVM element corresponds to the predicted label.


\subsection{Simulation and Results}

To demonstrate the ability of the VQSD for supervised classification, we performed classical simulations using PennyLane. The simulation uses the Iris flower dataset for ternary classification. The dataset consists of four attributes (a four-dimensional data point $\mathbf{x}_n$) and species (a class label $y_n$) of Iris flowers, and can be expressed as $\mathcal{D}=\{(\mathbf{x}_n,y_n)|\, \mathbf{x}_n\in\mathbb{R}^4, y_n\in\{0,1,2\}, n=1,\ldots, 150\}$.
Each class has the same number of data points (50 per class). 

\begin{figure}[t]
    \centering
    \includegraphics[width=0.45\textwidth]{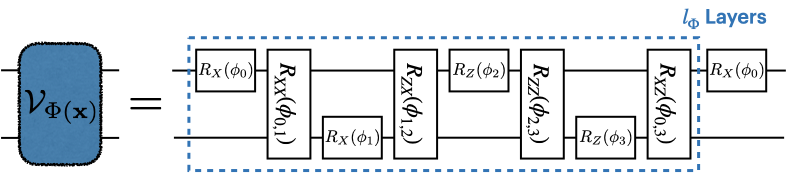}\\
    (a)\\ \vspace{0.3cm}
    \includegraphics[width=.42\textwidth]{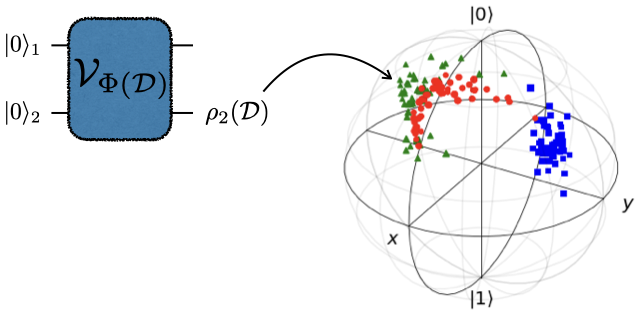}\\
    (b)
    \caption{\small Circuit diagrams for quantum feature map and Visualization of Iris flower dataset on Bloch sphere. 
    (a) A quantum feature map $\mathcal{V}_\Phi$ consists of single-qubit rotations and the Ising coupling gates. The order of gates in this map is determined by considering the commutation relation between them. 
    (b) The mixed states on the Bloch sphere are the reduced density matrices of the data, $\rho_2(\mathcal{D}) = \text{Tr}_1[\mathcal{V}_{\Phi(\mathcal{D})} |00\rangle\langle 00|\mathcal{V}_{\Phi(\mathcal{D})}^\dagger]$ with an encoding function Eq.\eqref{eq:invcoscos}. The blue square, red circle, and green triangle correspond to the data label 0, 1, and 2, respectively.}
    \label{fig:IrisBloch}
\end{figure}

\begin{figure*}[t!]
\centering
\begin{subfigure}{0.4\textwidth}
    \includegraphics[width=0.8\textwidth]{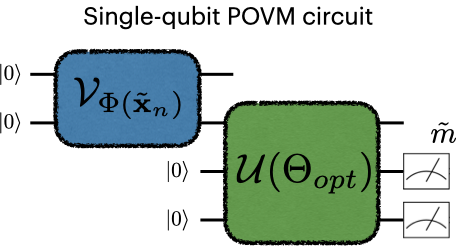}
\end{subfigure}
\hspace{10pt}
\begin{subfigure}{0.4\textwidth}
    \includegraphics[width=0.8\textwidth]{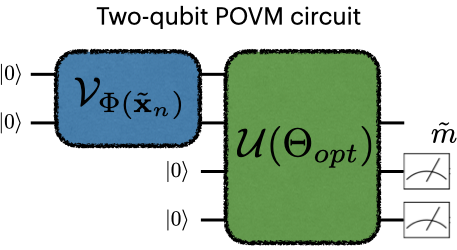}
\end{subfigure}
\\
\vspace{7pt}
\begin{subfigure}{0.4\textwidth}
    \includegraphics[width=\textwidth]{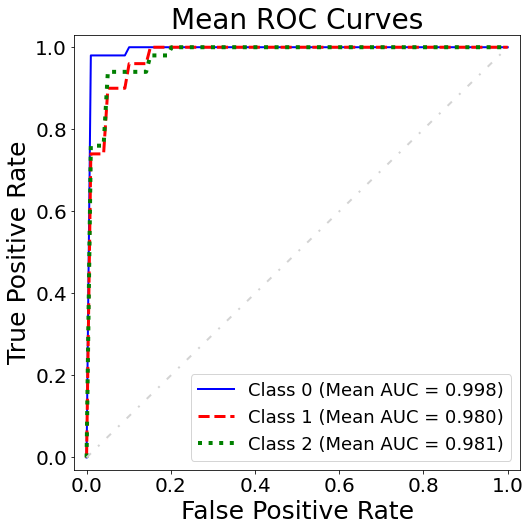}
    \caption{}
\end{subfigure}
\hspace{10pt}
\begin{subfigure}{0.4\textwidth}
    \includegraphics[width=\textwidth]{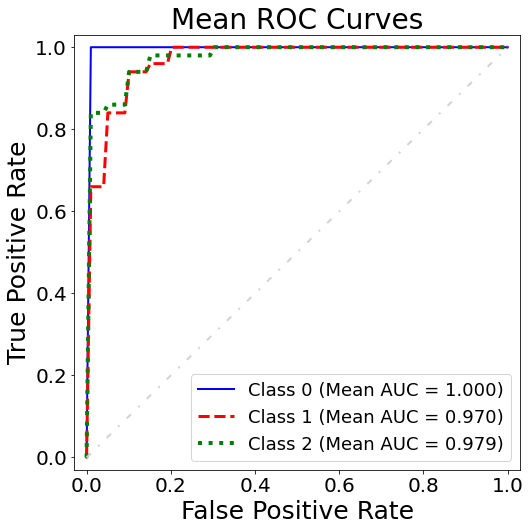}
    \caption{}
\end{subfigure}
\caption{{\small Receiver operating characteristic (ROC) analysis for Iris flower data classification by using VQSDs with different $n_T$, (a) the single-qubit ($n_T=1$) POVM circuit and (b) the two-qubit ($n_T=2$) POVM circuit. For each class, the mean ROC curves and mean AUCs averaged over the 5-fold cross-validation are presented in both quantum state discriminators. In each plot, the diagonal line corresponding to TPR=FPR is plotted as a guide to the eyes.
}}%
\label{fig:roc}
\end{figure*}


Applying quantum machine learning to classical data requires an initial step that encodes the classical data into a quantum state.
For this, various approaches exist, such as the amplitude encoding, angle encoding, Hamiltonian encoding, and trainable quantum encoding~\cite{SupervisedQML,lloyd2020quantum, araujo_divide-and-conquer_2021,9259210, PhysRevA.102.032420,araujo2021configurable,Havlicek2019,Blank_2022}.
In this simulation, we modify the Instantaneous quantum polynomial (IQP) encoding that utilizes Ising Hamiltoinan, motivated by Havl\'i\v{c}ek et al. \cite{Havlicek2019}, to map a four-dimensional data point into a two-qubit state.
The feature-encoding method suggested in Refs.~\cite{Havlicek2019,s42484-020-00020-y} maps classical data features to the interaction terms of the Ising Hamiltonian. This leads to using the single-qubit rotation gate $R_Z(\phi)=\exp(i\phi Z_i)$ and the two-qubit coupling gate $R_{ZZ}(\phi)=\exp(i\phi Z_iZ_j)$ to embed a two-dimensional input vector into two qubits. In addition to the $Z$ operators, we use $X$ operators because we wish to encode four data features into a two-qubit state $|\Psi_n\rangle=|\Phi(\mathbf{x}_n)\rangle = \mathcal{V}_{\Phi(\mathbf{x}_n)}|00\rangle$.
A quantum feature map is then defined as $\mathcal{V}_{\Phi}= (V_{\Phi})^{l_\Phi} e^{i\phi_1P_1}$, where 
\begin{align}
    V_{\Phi}= \prod^4_{\gamma=1} e^{\phi_{\gamma,\gamma+1}P_\gamma \cdot P_{\gamma+1}}e^{i\phi_\gamma P_\gamma}, 
\end{align}
$\phi$ is the coefficient that contains classical information, $l_\Phi$ is the number of layers, and $P_\gamma$ is the $\gamma$-th element of the set $\{X{\small\otimes}I, I{\small\otimes}X, Z{\small\otimes}I, I{\small\otimes}Z\}$. We do not insert Hadamard gates between encoding layers because each layer already contains non-commuting terms. The circuit diagram of $\mathcal{V}_\Phi$ is illustrated in in Fig.~\ref{fig:IrisBloch}(a).
In the quantum circuit for $\mathcal{V}_\Phi$, we encode data $\mathbf{x}_n$ into the coefficients $\phi$ by using the encoding function $\phi(\mathbf{x})$. 
The couplings in the quantum feature map $\mathcal{V}_\Phi$ necessitate the information on correlations among the attributes of a data point.
There are a variety of approaches to encode a classical data point $\mathbf{x}$ into the coefficients $\phi$.
We tested all encoding functions suggested in Refs~\cite{Havlicek2019,s42484-020-00020-y}, and found the following two types work the best for Iris data classification.
\begin{align}
\label{eq:invcoscos} \phi_j(\mathbf{x}) &= x_j, \;\;\; \phi_{i,j}(\mathbf{x}) = \frac{\pi}{3\cos(x_i)\cos(x_j)}\\
\label{eq:gaussian} \phi_j(\mathbf{x}) &= x_j, \;\;\; \phi_{i,j}(\mathbf{x}) = \exp\left(\frac{|x_i-x_j|^2}{8/\ln(\pi)}\right)
\end{align}
Thus, the simulation results in this section are of those carried out with the above encoding functions. 

Before encoding in the Hilbert space, let the Iris data set be preprocessed by rescaling each attribute into the range $[-\pi/5, \pi/5]$. This rescaling is empirically found to work well in combination with both encoding functions.
The rescaling mapping we use here is defined as 
\begin{equation}
x'_{n,\alpha} = \frac{\pi}{5}\frac{2x_{n,\alpha} - (x_{\min,\alpha} + x_{\max,\alpha})}{x_{\max,\alpha}-x_{\min,\alpha}},
\end{equation}
where $x_{\min,\alpha}:=\min_n x_{n,\alpha}$ and $x_{\max,\alpha}:=\max_n x_{n,\alpha}$ for each attribute $\alpha\in\{0,1,2,3\}$.
Next, the rescaled data point $\mathbf{x}_n'$ is embedded into the quantum state $|\Psi_n\rangle = \mathcal{V}_{\Phi(\mathbf{x}'_n)}|00\rangle$ using the quantum feature map $\mathcal{V}_\Phi$ and two encoding functions Eq.\eqref{eq:invcoscos} and Eq.\eqref{eq:gaussian}.
Here, the number of layers of the encoding circuit $\mathcal{V}_\Phi$ is empirically chosen as $l_\Phi=2$.

In conjunction with this quantum data encoding, we facilitate two different structures of the parameterized quantum circuit in the VQA: the single-qubit POVM circuit ($n_T=1$) and the two-qubit POVM circuit ($n_T=2$) whose diagrams are shown in Fig.~\ref{fig:roc}(a) and (b), respectively.
The encoding function Eq.\eqref{eq:invcoscos} is empirically selected for the former and Eq.\eqref{eq:gaussian} for the latter.
In both cases, measuring two ancilla qubits in the computational basis yields at most four outcomes of POVMs.

To assess the classification performance of the VQSD, we use the stratified 5-fold cross-validation method. 
Namely, we shuffle the Iris data set, split the data set in a 1:4 ratio of test to training data, and perform supervised classification on all five splits.
The classification performance of the VQSD is quantified by the average test performance of the five classification instances of the cross-validation process. Two metrics are used to determined the test performance in each instance: the accuracy and the area under the receiver operating characteristic (ROC) curve. The procedure for computing theses metrics in our algorithm is described in the following.

First, when an unlabeled data point $\tilde{\mathbf{x}}$ is given, the quantum state discriminator predicts the corresponding label $\tilde{y}=\tilde{m}(\tilde{\mathbf{x}})$ according to a decision rule. The decision rule chooses the label with the highest probability among all possible outcomes of POVMs, which is given by 
\begin{align}
    \tilde{m}(\tilde{\mathbf{x}}) &= \arg \max_m p(m|\tilde{\mathbf{x}}),
\end{align}
where $p(m|\tilde{\mathbf{x}}) = \langle\Phi(\tilde{\mathbf{x}})| \mathcal{E}_m(\Theta_{opt}) |\Phi(\tilde{\mathbf{x}})\rangle$ is the probability of obtaining the outcome $m$ from performing POVMs on $|\Phi(\tilde{\mathbf{x}})\rangle$.
For a test data set with known labels $\tilde{\mathcal{D}}=\{(\tilde{\mathbf{x}}_n,\tilde{y}_n)|\, \tilde{\mathbf{x}}_n\in\mathbb{R}^4, \tilde{y}_n\in\{0,1,2\}, \forall n\in\mathbb{N}\}$, one can examine how many labels matches the predictions of the quantum model by evaluating the accuracy $\sum_n\delta_{\tilde{m}(\tilde{\mathbf{x}}_n),\tilde{y}_n}/|\tilde{\mathcal{D}}|$. 
In the 5-fold cross-validation, the mean accuracy is 0.893 for the single-qubit POVM circuit and 0.920 for the two-qubit POVM circuit.

The second measure is the ROC curve and the area under the curve (AUC). Although this measure is designed for binary classification, we can apply it to multi-class classification through a one-vs-rest scheme.  
After formulating the multi-class classification to the one-vs-rest binary classification, the test data is assigned a label whose probability is higher than a predetermined threshold value. Then the true-positive rate and the false-positive rate can be evaluated at the threshold. The ROC curves are then drawn by shifting the threshold value. The farther the ROC curve from the diagonal line the better the performance of the classifier. Therefore, the performance of the classifier can be quantified by the area under the ROC curve (AUC), which ranges from 0.5 to 1, and the larger the AUC of ROC the better the performance of the classifier.
The mean ROC curves of the VQSD obtained from the 5-fold cross-validation are shown in Fig.~\ref{fig:roc}. There are three ROC curves in each plot corresponding to three different combination of the one-vs-rest binary classification. 
The VQSD classifiers with single-qubit and two-qubit POVM circuits both show excellent classification performance as the mean AUC ranges from 0.97 to 1 with an average of 0.985.

\section{Conclusions and Discussion}
\label{sec:conclusion}

We proposed the VQSD, a variational quantum algorithm that performs the minimum-error QSD. 
The proposed algorithm leverages the cosine-sine decomposition to design a parameterized POVM circuit, and utilizes a cost function derived from the theory of QSD to train the circuit.

The VQSD learns to solve the state discrimination tasks without prior knowledge of the target states. Our results indicate that this method is as effective as the semidefinite programming approach. We presented a multi-class classification algorithm based on the VQSD and showcase its efficacy using the Iris flower dataset. The numerical simulations conducted on this dataset produced an area under the receiver operating characteristic curve that ranged from 0.97 to 1, demonstrating its outstanding performance.


The VQSD offers several advantages over the classical solver for the SDP. Unlike the classical algorithm for solving the SDP, the VQSD does not require full information about the quantum states through quantum state tomography. Besides, the VQSD has a computational complexity of $O(2^{2n_T}l)$ {\small CNOT} gates, while the complexity of the classical SDP solver is $O(2^{\omega n_T}l^{\omega})$, where $\omega$ depends on the matrix multiplication algorithm and ranges from 2.37188 to 3. Additionally, classical optimization procedures require an extra step of finding the corresponding quantum circuit to implement QSD. These benefits make the VQSD more efficient than the classical SDP solver for calculating the probabilities from POVMs on quantum states and for discriminating multiple quantum states.

The VQSD can serve as a multi-class classifier in supervised machine learning by discriminating mixed states that encode data subject to classification. Since the quantum state can represent $O(2^{n_T})$ features~\cite{PhysRevLett.100.160501,araujo_divide-and-conquer_2021,9259210,araujo2021configurable,Blank_2022}, the VQSD circuit is efficient with respect to the number of data features. The VQSD classifies multi-class data sets by synchronizing data labels and measurement outcomes, requiring $\lceil \log_2 l\rceil$ ancillary qubits for $l$ data labels. Furthermore, existing quantum data compression techniques, such as the quantum convolutional neural network (QCNN)~\cite{cong_quantum_2019,maccormack_branching_2020,hur2021quantum} and the quantum autoencoder (QAE)~\cite{Romero_2017,PhysRevLett.124.130502} can reduce number of target qubits. This leads to a reduced size of the quantum circuit needed for multi-class classification, making implementation on NISQ devices more feasible. The use of the VQSD for multi-class classification in this manner represents a promising approach for addressing the challenges associated with implementing multi-class classification on quantum computing devices.

The performance of the VQSD can be improved by introducing an appropriate cost function. A known issue with the minimum-error quantum state discrimination strategy for more than two quantum states is that it can result in a zero-valued POVM, as discussed in~\cite{Bae_2015}. This problem can also occur in the implementation of the VQSD, as demonstrated in Fig.~\ref{eq:med}(b). Here, the optimal POVMs are given as $E_0=|0\rangle\langle 0|$, $E_1=|1\rangle\langle 1|$, and $E_2=0$, making it challenging to discriminate the state $|+\rangle$. As a result, the VQSD-based multi-class classifier may lose a data label. To address this issue, further research is needed to investigate other QSD strategies, such as maximum-confidence discrimination. Another avenue of exploration is to integrate the VQSD as the classifier in existing quantum machine learning algorithms, such as the QCNN, QAE, and the quantum generative adversarial network~\cite{PhysRevLett.121.040502,PhysRevA.98.012324,PhysRevApplied.16.024051}. As previously noted, the QCNN and the QAE can complement the VQSD by reducing the number of target qubits required for quantum state discrimination and, therefore, reducing the depth of the quantum circuit. Furthermore, exploring the application of the VQSD for unsupervised data clustering represents an intriguing direction for future research.

\section*{Acknowledgment}
This research was supported by the Yonsei University Research Fund of 2022 (2022-22-0124), by the National Research Foundation of Korea (2021M3E4A1038308, 2021M3H3A1038085, 2022M3H3A106307411, and 2022M3E4A1074591), and by the KIST Institutional Program (2E32241-23-010). We thank Israel Ferraz de Araujo for the insightful discussions on the subjects of gate decomposition and quantum circuit complexity.


\appendix

\section{Decomposition of the POVM circuit.}\label{append:decomp}
\,

\begin{figure}[h]
    \centering
    \includegraphics[width=.35\textwidth]{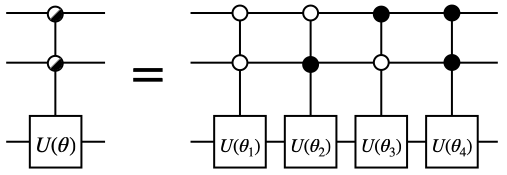}
    \caption{\small The uniform controlled gate \cite{Mottonen.2004,Bergholm.2005}. It is two-fold uniformly controlled gate $U(\boldsymbol{\theta})$ composed of four two-fold controlled $U$ gates, each with different angles $\theta_i$. The parameter vector $\boldsymbol{\theta}$ contains the angles $\theta_i$ for all $i$. The white control qubit represents that the unitary operation $R$ is applied only when the control qubit is the state $|0\rangle$.} 
    \label{fig:ucgate}
\end{figure}

We demonstrate the derivation of the circuit structure $\mathcal{U}$ in Fig.~\ref{fig:circ_decomp} based on the cosine-sine decomposition (CSD).
The CSD is closely related to the singular value decomposition (SVD) of a unitary operator. 
Let us consider an arbitrary $2^n$-by-$2^n$ unitary operator $V_n\in\text{U}(n)$ is divided into four blocks as depicted in the form of a 2-by-2 matrix.
The SVD is applied to each block, leading to the CSD and the circuit diagram of a unitary operator $V_n$ can be described as
\begin{center}
\includegraphics[width=.35\textwidth]{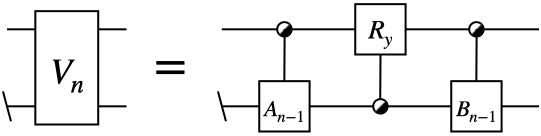}
\end{center}
\begin{align}\label{eq:csd}
    V_n = 
    \begin{pmatrix}
        B^{(0)}_{n-1} &  \rvline & \\
        \hline
        & \rvline & B^{(1)}_{n-1} 
    \end{pmatrix}
    \begin{pmatrix}
        C & \rvline & -S \\
        \hline
        S & \rvline & C
    \end{pmatrix}
    \begin{pmatrix}
        A^{(0)}_{n-1} & \rvline & \\
        \hline
        & \rvline & A^{(1)}_{n-1}
    \end{pmatrix}.
\end{align}
The two-fold uniformly controlled $A_{n-1}$ ($B_{n-1}$) gate corresponds to the first (third) decomposed matrix in \eqref{eq:csd} where $A^{(0)}_{n-1},A^{(1)}_{n-1}$ ($B^{(0)}_{n-1},B^{(1)}_{n-1}$) are left (right) unitary operators with dimensions of $2^{n-1}\times 2^{n-1}$ from the SVDs. 
The $(n-1)$-fold uniformly controlled $R_y$ gate matches the second decomposed matrix in Eq.\eqref{eq:csd} involving two different singular value matrices $C=diag(\cos{\varphi}_1, \cdots,\cos{\varphi}_{2^{n-1}})$ and $S=diag(\sin{\varphi}_1, \cdots,\sin{\varphi}_{2^{n-1}})$ for rotational angles $\varphi_i$ in $R_y$ gates.  
When one ancillary qubit is initially in the state $|0\rangle$, the CSD of $V_n$ can be simplified as
\begin{center}
    \includegraphics[width=.43\textwidth]{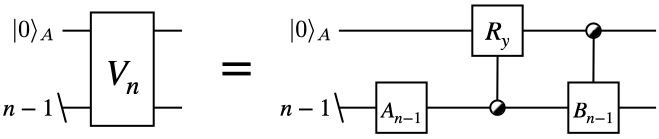}
\end{center}
\begin{align}\nonumber
    V_n &= 
    \begin{pmatrix}
        B^{(0)}_{n-1} & \rvline & \\
        \hline
        & \rvline & B^{(1)}_{n-1}
    \end{pmatrix}
    \begin{pmatrix}
        C & \rvline & -S \\
        \hline
        S & \rvline & C
    \end{pmatrix}
    \begin{pmatrix}
        A^{(0)}_{n-1} \\
        \hline
        \,
    \end{pmatrix}\\ 
    &= 
    \begin{pmatrix}
        B^{(0)}_{n-1} & \rvline & \\
        \hline
        & \rvline & B^{(1)}_{n-1}
    \end{pmatrix}
    \begin{pmatrix}
        C \\
        \hline
        S
    \end{pmatrix}
    A^{(0)}_{n-1}, \label{eq:csd_1a}
\end{align}
which contains $(n-1)$-qubit unitary operations $R_{n-1}$ without controls \cite{Iten.2016}.

Based on \eqref{eq:csd_1a}, we here demonstrate the circuit decomposition of $\mathcal{U}=V_n$ in Fig.~\ref{fig:circ_decomp}. 
When $n_A$ ancillary qubits initially set in the state $|\mathbf{0}\rangle_A$ are considered, one can decompose an arbitrary unitary operator $V_n$,
\begin{center}
\includegraphics[width=.38\textwidth]{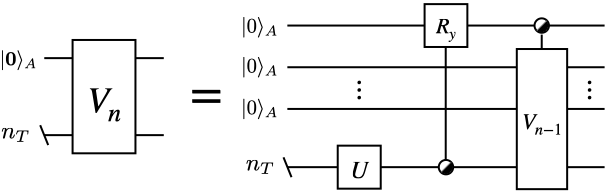}
\end{center}
\begin{align}\label{eq:csd_na}
    V_n
    = 
    \begin{pmatrix}
        V_{n-1}^{(0)} & \rvline & \\
        \hline
        & \rvline & V_{n-1}^{(1)}
    \end{pmatrix}
    \begin{pmatrix}
        C \\
	\\ 
        \hline
        S \\ 
        \\ 
    \end{pmatrix}
    U, 
\end{align}
where $U$ is a $2^{n_T}$-by-$2^{n_T}$ unitary operator corresponding to the right unitary matrix from performing the SVD on $V_n$,  
$C$ (or $S$) is the $2^{n_T}$-by-$2^{n_T}$ diagonal matrix satisfying $C_{ii}=\cos \varphi_i$ (or $S_{ii}=\sin\varphi_i$), 
and $V^{(z_1)}_{n-1}$ for any $z_1\in\{0,1\}$ is a unitary operator of size $2^{n-1} \times 2^{n-1}$ with $n=n_T+n_A$.
$V^{(z_1)}_{n-1}$ specifically consists of $2^{n_T}$ columns determined by the left-singular vectors of $V_n$ and arbitrary $2^{n_A-1}$ columns.
Each block in the second decomposed matrix of \eqref{eq:csd_na} contains $2^{n_A-1}-1$ zero matrices of dimension $2^{n_T}$ \cite{Iten.2017}.
By using \eqref{eq:csd_na} recursively, a unitary operator $V_n$ can be decomposed until $V_{n_T}$ is obtained,
\begin{center}
\includegraphics[width=.43\textwidth]{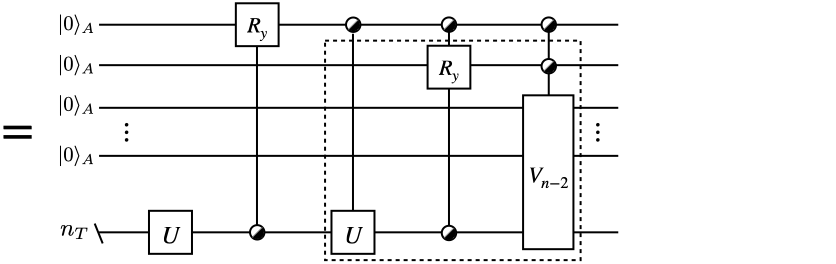}
\includegraphics[width=.43\textwidth]{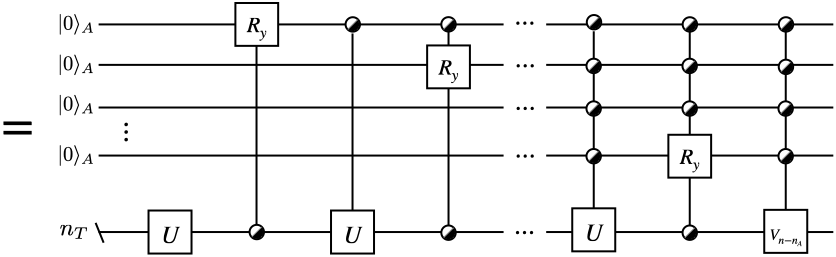}
\end{center}
Thus, this circuit structures are exactly the same as the ones in Fig.~\ref{fig:circ_decomp}.
when the $n_A$-fold uniformly controlled $V_{n_T}$ gate at the end is omitted, which does not lose the generality of performing POVMs on $n_T$-qubit states.
Moreover, the circuit can be more simplified by commuting the control and the measurement, thereby reducing the number of the {\small CNOT} gate and considering it only for the gate $U$ \cite{Andersson.2008, Iten.2017, chen_qumi2020}.

\section{Semidefinite programming for minimum-error QSD}\label{append:sdp}

Semidefinite programming (SDP) provides numerical solutions in numerous applications of quantum information theory \cite{watrous_2018}. For instance, SDP is used to compute monotones quantifying quantum resources in convex quantum resource theories \cite{Chitambar2019} such as entanglement\cite{Doherty2002}, steering\cite{Cavalcanti_2016} and coherence\cite{Streltsov2017}, and also to calculate the minimum error discrimination in discrimination tasks \cite{Bae_2015, Gilchrist2005}.

The (primal) SDP is formulated as follows: 
\begin{align}\label{eq:sdp}
    \text{Minimize } \;\;\;\;&\text{Tr}[\mathcal{C}X],\\
    \text{subject to }  \;\;\;\; &\text{Tr}[\mathcal{A}_jX]\le b_j \nonumber \text{ for } j=0,...,k-1,\\
    & X \ge 0, \nonumber
\end{align}
where $h\times h$ Hermitian matrices $\mathcal{C}, \mathcal{A}_j$ and real number $b_j$ are given for all integer $j$. The objective function is minimized over the positive semidefinite matrix $X$ of dimension $h\times h$.
The recent method for solving SPD has been known as runtime $O(k(k^2+h^\omega+ks)\log(1/\epsilon))$ where the constant $\omega$ is the exponent of a matrix multiplication algorithm, $s$ is the sparsity of given matrices $(\mathcal{C}, \mathcal{A}_0,\cdots,\mathcal{A}_{k-1}$), and $\epsilon$ is the accuracy of the solution \cite{Lee.2015qd}.

The minimum-error discrimination task in Section~\ref{sec:med} can be viewed in the framework of SDP.
The error probability Eq.\eqref{eq:med} is equivalent to the objective function of SDP by considering $\mathcal{C}:=\bigoplus_{j=0}^{l-1} -q_j\rho_j$ and $X=\bigoplus_{j=0}^{l-1} E_j$.
Also, the completeness of the POVM can be assigned to an equality constraint $\text{Tr}[X]=h/l$ in \eqref{eq:sdp}.
If the quantum system is defined on the $d$-dimensional Hilbert space, the dimension $h$ is given as the multiplication of $d$ and $l$. 
It is remarkable that computing SDP has polynomial time complexity with the dimension of the Hilbert space $d$ but exponential one with the number of target qubits $n_T$ in this work.
Moreover, in order to construct SDPs, the exact form of states should be specified. Namely, quantum state tomography (QST) is required. On the other hand, QST is not involved in our framework because parameter optimization is performed without state information. As a result, this property allows us to apply our VQAs to find the optimal POVMs in machine learning straightforwardly.



\end{document}